\documentclass[preprint]{elsarticle}

\usepackage{geometry}
\biboptions{authoryear}
\usepackage{multicol}
\usepackage{url}

\usepackage{xcolor}
\usepackage{scalerel}
\usepackage{tikz} \usetikzlibrary{svg.path}
\definecolor{orcidlogocol}{HTML}{A6CE39}
\usepackage{hyperref, doi}

\makeatletter
\def\ps@pprintTitle{%
 \let\@oddhead\@empty
 \let\@evenhead\@empty
 \def\@oddfoot{\centerline{\today}}%
 \let\@evenfoot\@oddfoot}
\makeatother

\usepackage{csquotes, amsmath, amssymb, bm, bbm}

\newcommand{\myxi}{\bm{\xi}}
\newcommand{\myD}{\textbf{D}}
\newcommand{\myO}{\textbf{O}}
\newcommand{\myT}{\mathcal{T}}
\newcommand{\myA}{\mathcal{A}}
\newcommand{\myDT}{\myD_{\myT}}
\newcommand{\myDA}[1]{\myD_{\myA_{#1}}}
\newcommand{\myxiT}{(\myxi_t)_{t \in \myT}}
\newcommand{\myxiA}{(\hat{\myxi}_{\myA(t)})_{t \in \myT}}
\newcommand{\myxiAO}{(\hat{\myxi}_{\myA_0(t)})_{t \in \myT}}
\newcommand{\myxiAI}{(\hat{\myxi}_{\myA_1(t)})_{t \in \myT}}
\newcommand{\gen}{\text{gen}}
\newcommand{\tr}{\text{tr}}
\newcommand{\ch}{\text{ch}}
\newcommand{\sto}{\text{sto}}
\newcommand{\capgen}{\text{cap}^{\gen}}
\newcommand{\captrans}{\text{cap}^{\tr}}
\newcommand{\capsto}{\text{cap}^{\sto}}

\newcommand{\ESMplan}{\Phi_{\text{plan}}}
\newcommand{\ESMoperate}{\Phi_{\text{operate}}}

\usepackage{graphicx} \graphicspath{{figures/}}

\begin{document}

\begin{frontmatter}
\title{Reducing climate risk in energy system planning: a posteriori time series aggregation for models with storage}
\author[imperial]{Adriaan P Hilbers\corref{mycorrespondingauthor}}
\cortext[mycorrespondingauthor]{Corresponding author}
\ead{a.hilbers17@imperial.ac.uk}
\author[reading]{David J Brayshaw}
\author[imperial]{Axel Gandy}
\address[imperial]{Department of Mathematics, Imperial College London}
\address[reading]{Department of Meteorology, University of Reading}

\begin{abstract}
  \hspace{\parindent}The growth in variable renewables such as solar and wind is increasing the impact of climate uncertainty in energy system planning. Addressing this ideally requires high-resolution time series spanning at least a few decades. However, solving capacity expansion planning models across such datasets often requires too much computing time or memory.

  To reduce computational cost, users often employ \textit{time series aggregation} to compress demand and weather time series into a smaller number of time steps. Methods are usually \textit{a priori}, employing information about the input time series only. Recent studies highlight the limitations of this approach, since reducing statistical error metrics on input time series does not in general lead to more accurate model outputs. Furthermore, many aggregation schemes are unsuitable for models with storage since they distort chronology.

  In this paper, we introduce \textit{a posteriori} time series aggregation schemes for models with storage. Our methods adapt to the underlying energy system model; aggregation may differ in systems with different technologies or topologies even with the same time series inputs. Furthermore, they preserve chronology and hence allow modelling of storage technologies.

  We investigate a number of approaches. We find that \textit{a posteriori} methods can perform better than \textit{a priori} ones, primarily through a systematic identification and preservation of relevant extreme events. We hope that these tools render long demand and weather time series more manageable in capacity expansion planning studies. We make our models, data, and code publicly available.
\end{abstract}


\end{frontmatter}

\newpage
\section{Introduction}
\label{sec:introduction}

\subsection{Capacity expansion planning models}
\label{sec:introduction:espms}

The growth of variable renewables such as solar and wind has created new computational challenges in optimisation-based energy system planning. This is because accurate representation of such technologies' variability requires both a high spatiotemporal resolution \citep{Kools2016, Poncelet2016, collins_2017} and long simulation lengths \citep{Bloomfield2016, staffel_2018, zeyringer_2018, collins_2018, Hilbers2019, Kumler2019, Bryce2018, Shaner2018, Hilbers2021}. This leads to high computational costs, since algorithms to solve the associated optimisation problems scale quickly (often exponentially) in the number of time steps \citep{Cao2019, Goderbauer2019}. It also hampers the study of climate impacts, since the use of climate model data --- typically various multi-year samples from an ensemble of simulations --- is currently unfeasible \citep{Bloomfield2020, Craig2022}.

In this paper, we consider \textit{capacity expansion planning} models, used to inform investments into energy infrastructure \citep{pfenninger_2014}. They determine the system design\footnote{In this paper, we consider \textit{build-from-scratch} models that determine the full system design. Ideas generalise naturally to investments into existing systems.} that minimises the sum of install and subsequent operation costs given a sample of demand and weather data \citep{Koltsaklis2018}. We view them as functions $\ESMplan$ from a demand and weather time series $\myxiT$ to the associated optimal system design (installed capacities of generation, transmission and storage technologies) $\myDT$:
\begin{equation}
  \label{eq:esm_plan}
  \myD_{\myT} = \ESMplan(\myxiT).
\end{equation}
The vector $\myxi_t$ contains time series values in period $t$. For example, for daily periods with hourly demand levels and wind speeds,
\begin{equation}
  \myxi_t = [
  d_{t,1}, \ \cdots, \ d_{t,24}, \
  w_{t,1}, \ \cdots, \ w_{t,24}
  ]
\end{equation}
where $d_{t,i}$ and $w_{t, i}$ are the demand and wind speed respectively in the $i$th hour of day $t$.

Each planning problem has an associated \textit{operation} problem (sometimes called the \textit{production cost model}), in which we fix the system design $\myD$ and optimise the system operation:
\begin{equation}
  \label{eq:esm_operate}
  (\myO_t)_{t \in \myT} = \ESMoperate(\myxiT \ | \ \myD)
\end{equation}
where $\myO_t$ contains the \textit{operational variables} in period $t$: generation, transmission and storage (dis)charge decisions.

\subsection{Time series aggregation}
\label{sec:introducion:time_series_aggregation}

\begin{figure}
  \centering
  \includegraphics[scale=0.8, trim=10 327 10 10, clip]{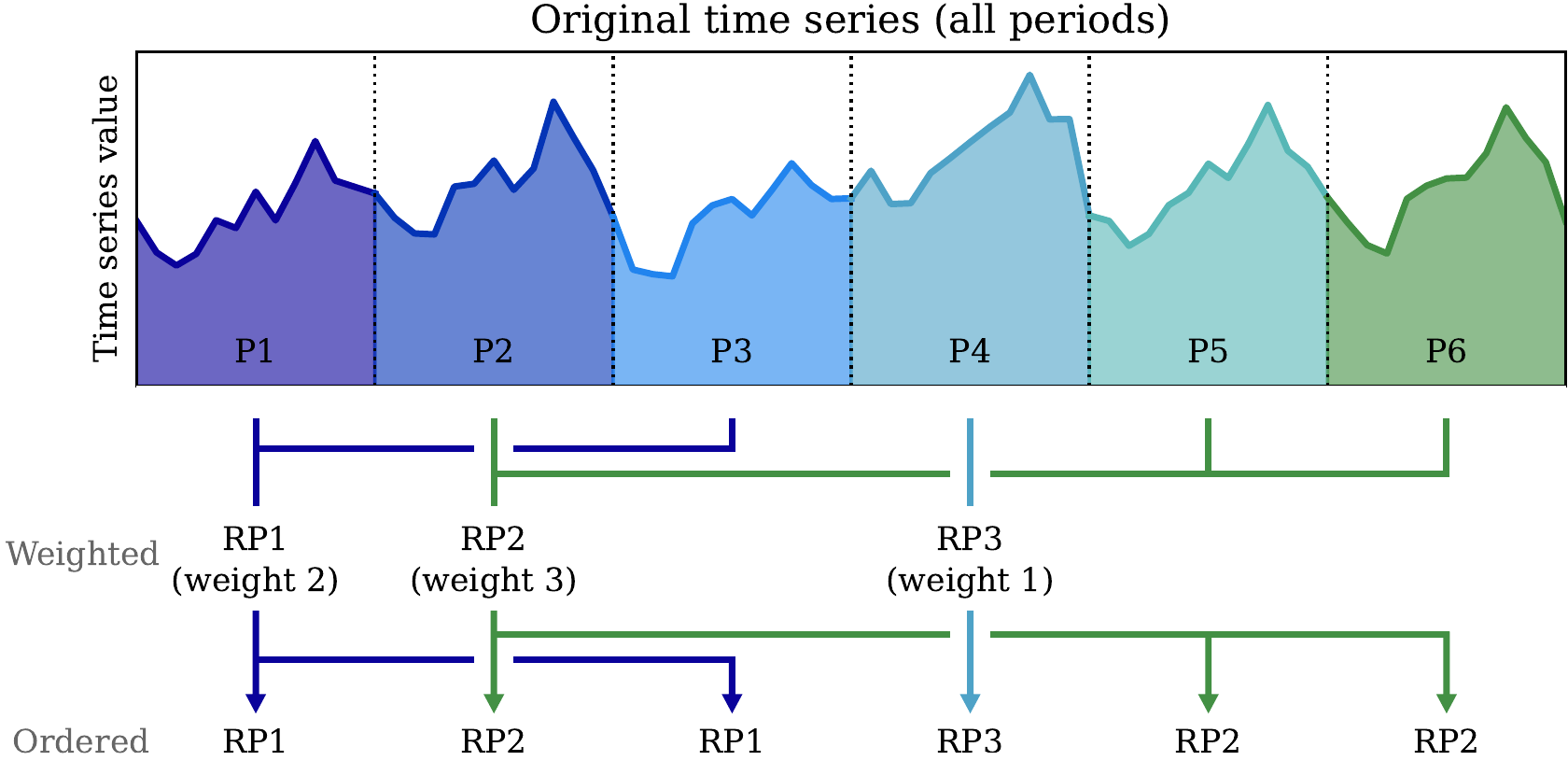}
  \caption{Time series aggregation from six periods (P1-6) to three \textit{representative periods} (RP1-3), either \textit{weighted} (appearing once, weighted by number of occurrences in full time series) or \textit{ordered} (in same order as full time series).}
  \label{fig:aggregation}
  \vspace{3em}
  \includegraphics[scale=0.8, trim=10 160 10 275, clip]{figures/ts_aggregation.pdf}
  \caption{Time series after aggregation into weighted representative days.}
  \label{fig:ts_aggregated}
  \vspace{3em}
  \includegraphics[scale=0.8, trim=10 10 10 425, clip]{figures/ts_aggregation.pdf}
  \caption{Storage levels after aggregation into ordered representative days and decomposition into \textit{inter-period} (level at start of period) and \textit{intra-period} (change compared to start of period) levels. Intra-period contributions are equal in each replication of the same representative period.}
  \label{fig:storage_aggregated}
\end{figure}

\textit{Time series aggregation}, as reviewed by \citet{hoffmann_2020} and \citet{Teichgraeber2022}, creates compressed time series that planning models are subsequently solved across. Many approaches create a smaller set of \textit{representative periods} as shown in Figure \ref{fig:aggregation}. An example mapping $\myA$, from six periods to three, is:
\begin{align}
  \myxiT = (\myxi_1 \quad \myxi_2 \quad &\myxi_3 \quad \myxi_4 \quad \myxi_5 \quad \myxi_6) \nonumber \\
  &\downarrow \\
  \myxiA = (\hat{\myxi}_1 \quad \hat{\myxi}_3 \quad &\hat{\myxi}_1 \quad \hat{\myxi}_2 \quad \hat{\myxi}_3 \quad \hat{\myxi}_3). \nonumber
\end{align}
The number of unique representative periods $\{\myA(t) \ | \ t \in \myT\}$ is usually significantly smaller than the number of original periods $\{t \ | \ t \in \myT\}$. The mapping $\myA$ can be determined in various ways, such as choosing days from each season \citep{Welsch2012}, minimising the deviation of load duration curves \citep{Sisternes2013, poncelet_2017} or clustering vectors of each period's time series values \citep{nahmmaccher_2016, poncelet_2017, hartel_2017, pfenninger_2017, Kotzur2018, Kittel2022}.

Without constraints linking periods, time series aggregation reduces computational cost since operational decision variables appear once per \textit{representative} period, weighted in the objective function by its number of occurrences \Citep{Merrick2016}. These are \textit{weighted} representative periods in Figures \ref{fig:aggregation} and \ref{fig:ts_aggregated}. We discuss linked periods in Section \ref{sec:introduction:inter_period_links}.

Relevant extreme events --- e.g.\ with high demand but low renewable output --- disproportionately drive accurate estimates of optimal design or cost \citep{Teichgraeber2020}. This sometimes motivates heuristic adjustments such as including the maximum demand or mininum renewable potential day \citep{pfenninger_2017, Kotzur2018}. However, such approaches may fail to identify the extremes relevant to the particular model. For example, peak demand may not require peak capacity if there is ample renewable generation or if stored or imported energy is available.

\subsection{Inter-period links and storage}
\label{sec:introduction:inter_period_links}

Constraints linking periods, such as storage, complicate time series aggregation since they require chronology of representative periods to be preserved. A number of solutions have been proposed; they include merging only periods that are adjacent chronologically \citep{Pineda2018, Tso2020, DeGuibert2020}, aggregating periods from different parts of the year separately \citep{Welsch2012, Samsatli2015, Timmerman2017}, and \textit{linking} storage levels between representative periods \citep{Gabrielli2018, tejada_2018, Kotzur2018a, VanderHeijde2019, Novo2022}.

Another complication is that aggregating time series inputs no longer automatically reduces the number of decision variables, since absolute storage levels may differ in replications of the same representative period (Figure \ref{fig:storage_aggregated}). For this reason, modellers often assume storage (dis)charge decisions are identical in each replication \citep{Gabrielli2018, tejada_2018, VanderHeijde2019}. \cite{Kotzur2018a} exploit this assumption by decomposing storage levels into \textit{inter-period} (level at start of period, one per original period) and \textit{intra-period} (deviation from start of period, one per representative period) contributions. Figure \ref{fig:storage_aggregated} shows this decomposition and how \textit{ordered} representative days with linked storage levels preserve chronology. For a detailed discussion, see \citep{Gonzato2021}.

\subsection{\textit{A posteriori} methods}
\label{sec:introduction:adaptive_methods}

Most time series aggregation schemes are what \cite{hoffmann_2020} call \textit{a priori}: they use information about the input time series only, creating identical aggregation for any model with the same time series inputs irrespective of technologies or topology. This has been criticised by \cite{Wogrin2022}, since reducing error metrics on time series inputs alone does not necessarily improve estimates of model outputs, i.e.\ optimal system design or cost \citep{Gonzato2021, Teichgraeber2022}.

\textit{A posteriori} (also known as \textit{adaptive}) methods use information about the underlying energy system model to tailor aggregation. For example, \citet{sun_2019} and \citet{Zhang2022} cluster vectors of planning model outputs (run on each individual day) instead of the time series itself. \citet{Bahl2018} and \citet{Teichgraeber2021} alternate between a planning model on aggregated data and an operation model on the full time series to iteratively identify and include days with unmet demand; this ensures design estimates have adequate generation capacity for such events. \cite{Hilbers2020} identify and include system-relevant extreme events using their generation cost, also calculated using an operation model. \citet{Li2022} combines elements of both such approaches.

\section{This paper's contribution}
\label{sec:contribution}

In this paper, we introduce \textit{a posteriori} time series aggregation schemes for capacity expansion planning models with storage. These schemes (1) tailor aggregation to the underlying energy system model and (2) preserve chronology, allowing the representation of long-term storage patterns.  To our best knowledge, we are the first to combine these approaches. We make our models, time series data and code available at \doi{10.5281/zenodo.7178301}.

We introduce a general aggregation framework that uses a model's operational variables --- generation, transmission and storage patterns --- to improve aggregation while maintaining chronology. It builds on work by \citet{Hilbers2020} and \citet{Teichgraeber2021} --- which uses operational variables in aggregation for models without storage --- and the chronology-preserving aggregation introduced by \citet{Kotzur2018a}.

These methods fill a research gap. From a modelling perspective, recent studies indicate that \textit{a priori} aggregation using time series inputs alone may lead to significant errors (Section \ref{sec:introduction:adaptive_methods}), while accurately representing storage technologies becomes essential as their role grows. From an applied perspective, our methods allow the consideration of multi-year samples in planning models at significantly lower error than current approaches; this is important for robust decision-making under climate uncertainty as discussed in Section \ref{sec:introduction:espms}. They also allow the use of climate model data --- typically long time series produced by an ensemble of climate models --- in planning studies.

This paper is structured as follows. Section \ref{sec:method} introduces the method, including the intuition behind its machinery. Section \ref{sec:siss:simulations} provides a case study application. In Section \ref{sec:conclusion_implications}, we discussion conclusions, implications and possible extensions.

\section{Methods}
\label{sec:method}

\subsection{Overview and intuition}
\label{sec:method:overview}

\begin{figure}
  \centering
  \includegraphics[scale=0.8, trim=10 10 10 0]{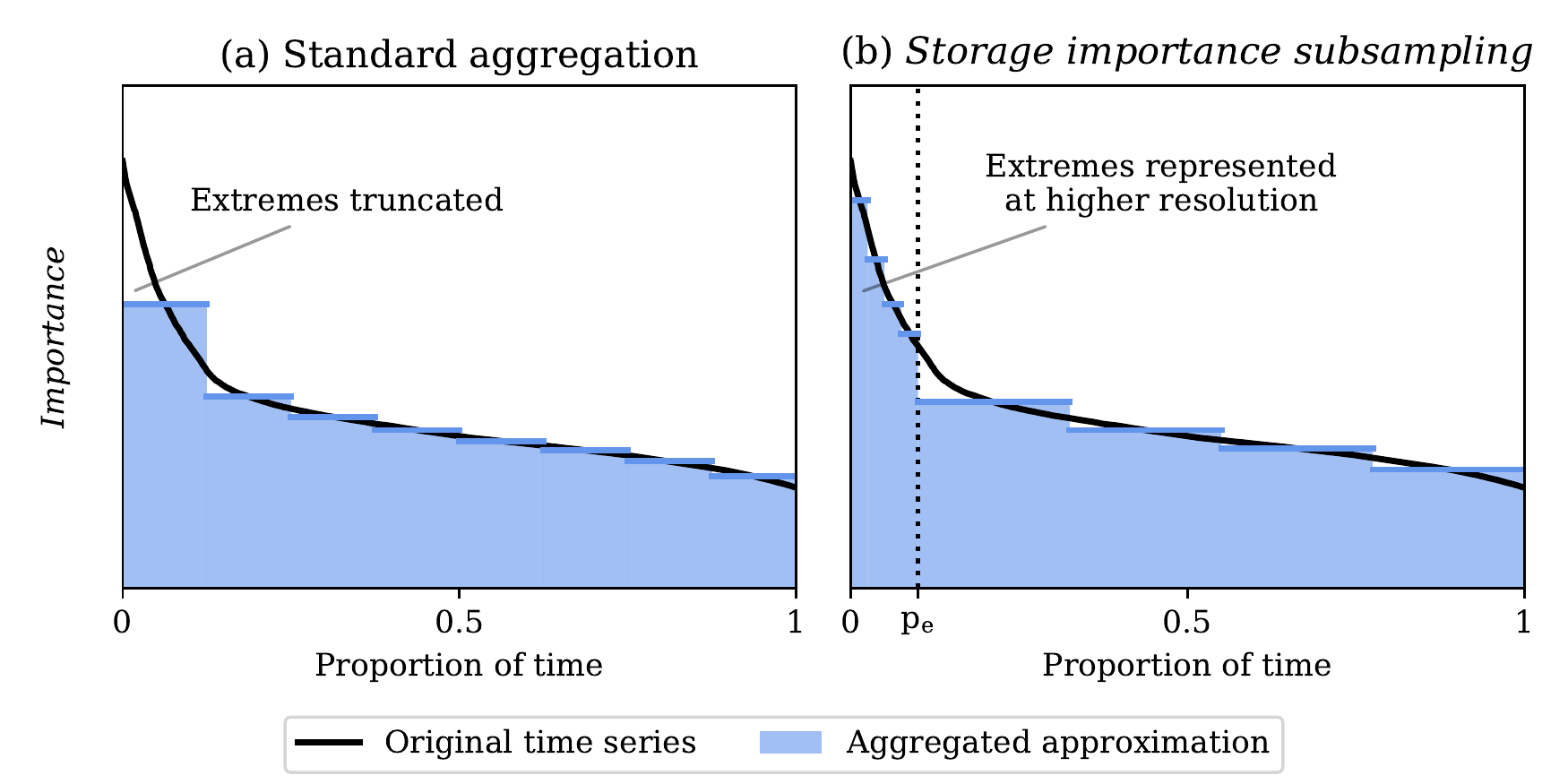}
  \caption{Impacts of time series aggregation. Black line: \textit{duration curve} (values plotted from highest to lowest) of the \textit{importance} (e.g.\ generation cost or electricity price), which identifies extreme events. (a) Standard aggregation truncates extremes. (b) Our method preserves extremes at higher resolution.}
  \label{fig:ldc}
\end{figure}

Consider two hypothetical planning models A and B. They take the same time series inputs, but A allows only fossil-fuel technologies, while B contains primarily variable renewables. \textit{A priori} aggregation leads to the same representative periods for both models, even though weather variables are unimportant for model A but very important for model B.

Our framework uses a model's \textit{operational variables} --- generation, transmission and storage patterns --- to customise time series aggregation. We can do this in two ways. The first is to model a selection of relevant extreme events --- as identified by an \textit{importance} function such as generation cost, electricity price or unmet demand --- at higher resolution (Figure \ref{fig:ldc}, see Section \ref{sec:method:remarks} for discussion). The second is by using operational variables when clustering. For example, concatenating storage (dis)charge decisions to the vectors we cluster encourages periods with similar storage patterns to be mapped to the same representative.

Unlike time series inputs, operational variables are not available \textit{a priori}. For example, we do not know a model's generation levels before simulations. We hence propose a two-stage approach. We determine a first-stage optimal design estimate $\myDA{0}$ using \textit{a priori} aggregation. We then calculate operational variables via an operational model (equation (\ref{eq:esm_operate})) across the full time series given $\myDA{0}$. These variables are used in a second planning model with \textit{a posteriori} aggregation.

\subsection{Framework: \textit{storage importance subsampling}}
\label{sec:method:steps}

Suppose we have a planning model $\ESMplan$ and want to estimate the optimal design $\myDT = \ESMplan(\myxiT)$ across a long sample $\myxiT$ of demand and weather data. We estimate it by $\myDA{1}$, determined from the following algorithm.

\vspace*{1em}
\noindent \begin{minipage}[c]{0.85\linewidth}
  \noindent \hrulefill \\
  \textbf{Inputs}:
  \begin{itemize}
    \vspace*{-0.7em}
    \setlength\itemsep{-0.3em}
  \item $\myxiT$: demand and weather time series, length $n_{\myT}$ periods
  \item $n_{\myA}$: number of unique representative periods to aggregate into
  \item $p_e \in [0, 1]$: proportion of periods in $\myT$ considered ``extreme''
  \item $\text{IMP}$: real-valued \textit{importance} function of operational variables $\myO_t$
  \end{itemize}
  \textbf{Steps}:
  \begin{enumerate}
    \vspace*{0.0em}
    \setlength\itemsep{0.7em}
  \item Get preliminary optimal design estimate $\myDA{0}$:
    \begin{enumerate}
      \vspace*{0.0em}
      \setlength\itemsep{0.3em}
    \item Aggregate $\myT$ into $\myxiAO$, $n_{\myA}$ unique representative periods, using \textit{a priori} scheme.
    \item Solve planning problem:
      \vspace*{-0.7em}
      \begin{equation}
        \myDA{0} = \ESMplan(\myxiAO).
        \vspace*{-0.7em}
      \end{equation}
    \end{enumerate}
  \item Create \textit{importance subsample}:
    \begin{enumerate}
      \vspace*{0.0em}
      \setlength\itemsep{0.3em}
    \item  Determine system operation across full time series:
      \vspace*{-0.7em}
      \begin{equation}
        (\myO_t)_{t \in \myT} = \ESMoperate(\myxiT \ | \ \myDA{0}).
        \vspace*{-0.7em}
      \end{equation}
    \item Calculate \textit{importance} of each period:
      \vspace*{-0.7em}
      \begin{equation}
        \text{imp}_t = \text{IMP}(\myO_t) \quad \text{for all} \ t \in \myT.
        \vspace*{-0.7em}
      \end{equation}
    \item Partition time series $\myT$ into:
      \begin{itemize}
      \item $\myT_e$: $p_e n_{\myT}$ ``extreme'' periods (with highest \textit{importance})
      \item $\myT_r$: $(1 - p_e) n_{\myT}$ ``regular'' periods (those remaining).
      \end{itemize}
    \item Aggregate into $\myxiAI$, $n_{\myA}$ unique representative periods, with:
      \begin{itemize}
      \item $\frac{n_{\myA}}{2}$ ``extreme'' representative periods aggregated from $\myT_e$
      \item $\frac{n_{\myA}}{2}$ ``regular'' representative periods aggregated from $\myT_r$.
      \end{itemize}
      Aggregate using input time series and/or operational variables.
    \end{enumerate}
  \item Get final optimal design estimate $\myDA{1}$:
    \vspace*{-0.7em}
    \begin{equation}
      \myDA{1} = \ESMplan(\myxiAI).
      \vspace*{-0.7em}
    \end{equation}
  \end{enumerate}
  \textbf{Output}: $\myDA{1}$: optimal design estimate
  \vspace*{-0.3em} \\
  \hspace*{-0.3em} \noindent \hrulefill
\end{minipage}

\subsection{Remarks}
\label{sec:method:remarks}

The method above is an \textit{a posteriori} scheme that tailors aggregation to the underlying planning model using (estimated) operational variables. An \textit{a priori} scheme would finish after step 1 and return $\myDA{0}$. We instead use $\myDA{0}$ to construct a new aggregation $\myA_1$ for a second, hopefully more accurate, optimal design estimate $\myDA{1}$. We preserve chronology for storage technologies using ordered and linked representative days as discussed in Section \ref{sec:introduction:inter_period_links} and Figures \ref{fig:aggregation}-\ref{fig:storage_aggregated}.

The \textit{importance} function $\text{IMP}$ in step 2(b) identifies relevant extreme events. It is one-dimensional to allow stratification into ``extreme'' and ``regular'' periods in step 2(c). It should identify events that require peak generation, transmission and storage capacities. In our case study (Section \ref{sec:siss:simulations}), we examine two candidates. The first is the generation cost, which naturally identifies extremes since expensive measures (e.g.\ peaking plants or load curtailment) are used only in settings where there are otherwise supply shortages. The second is unserved energy, which occurs only at times of insufficient supply. These are roughly those used by \citet{Hilbers2020} and \citet{Teichgraeber2021} respectively for models without storage. Expert knowledge can motivate others, e.g.\ electricity price.

Using this framework requires a number of choices. One is the \textit{importance} function discussed above. Another is $p_e$, the proportion of periods in the full time series $\myT$ considered extreme. This is a trade-off; a larger $p_e$ value (with $n_{\myA}$ fixed) means more periods are considered extreme, but are modelled at a lower resolution each (see Figure \ref{fig:ldc}). For simplicity, we use equal numbers of representative periods ($\frac{n_{\myA}}{2}$) for both ``extreme'' and ``regular'' regions. We must also specify the aggregation used in steps 1(a) and 2(d); options include e.g.\ $k$-means/medoids or hierarchical clustering.

\section{Simulation studies}
\label{sec:siss:simulations}

\subsection{Overview}
\label{sec:simulations:overview}

\begin{figure}
  \small
  \begin{tabular}{c}
    \includegraphics[scale=0.75, trim=10 10 10 10, clip]{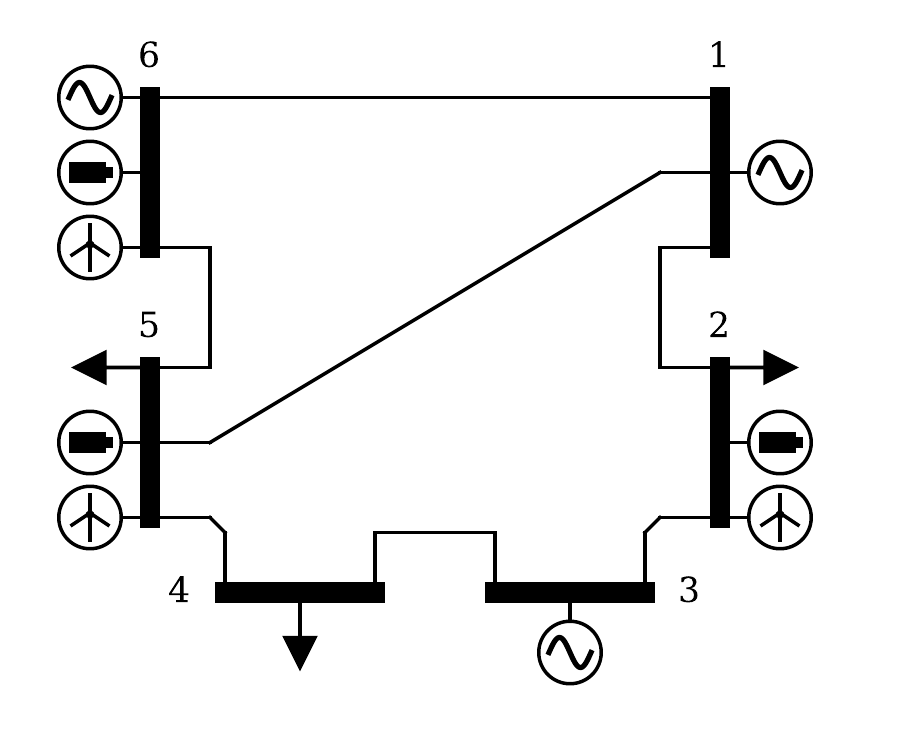}
  \end{tabular}
  \setlength{\tabcolsep}{0.4em}
  \begin{tabular}{r l l}
    region & technologies & time series \\ \hline
    1 & baseload, peaking & --- \\ [0.6em]
    2 & wind, storage & demand, wind (DE) \\ [0.6em]
    3 & baseload, peaking & --- \\ [0.6em]
    4 & --- & demand (FR) \\ [0.6em]
    5 & wind, storage & demand, wind (UK) \\ [0.6em]
    6 & baseload, peaking, & wind (ES) \\
      & wind, storage & \\ \hline
\end{tabular}
  \caption{Planning model topology. Demand and generation/storage technologies are distributed across six regions, linked by seven transmission lines. Regions 2, 4, 5 and 6 use time series data from Germany (DE), France (FR), the United Kingdom (UK) and Spain (ES) respectively.}
  \label{fig:model_diagram}
\end{figure}

In this section we examine the performance of a number of schemes, both \textit{a priori} and \textit{a posteriori} (in the framework of Section \ref{sec:method}) on an example energy system planning model. We conduct two experiments. The first is a validation exercise on a relatively short time series; we calculate the ``true'' (non-aggregated) optimal design $\myDT$ and compare it with aggregated estimates $\myDA{}$. The second uses a longer time series, for which calculating $\myDT$ directly is unfeasible; in this case we examine computational costs. In both experiments, we calculate \textit{unserved energy} --- demand unable to be met by a system with design $\myDA{}$ --- across the full time series.

We run experiments across the six-region planning model illustrated in Figure \ref{fig:model_diagram}. It determines the generation (baseload, peaking and wind), transmission and storage capacities that minimise the sum of install and operation costs given hourly demand levels and wind \textit{generation potentials} (generation as a fraction of rated capacity, also called \textit{capacity factors}). The model is based on a renewable version of the \textit{IEEE six-bus} system introduced by \citet{kamalinia_2011} and \citet{Hilbers2020}. For details, see \ref{app:psm}.

This section is structured as follows. Section \ref{sec:simulations:setup} describes our time series aggregation schemes. Sections \ref{sec:simulations:validation} and \ref{sec:simulations:example} present results from the validation study on short base time series and the example exercise on longer ones respectively. Section \ref{sec:simulations:discussion} discusses results and their implications.

\subsection{Setup}
\label{sec:simulations:setup}

\begin{table}
  \centering
  \small
  \setlength{\tabcolsep}{0.1em}
  \begin{tabular}{r p{31em}}
    \hline
    \textit{Method A (a priori)}: & Use cluster mean as representative day. \\
    \textit{Method B (a priori)}: & Use cluster medoid (closest real day to mean) as representative day. \\
    \textit{Method C (a priori)}: & Medoid repr., include maximum demand and mininum wind days. \\
    \textit{Method D (a posteriori)}: & Medoid repr., model days with unserved energy at higher resolution. \\
    \textit{Method E (a posteriori)}: & Medoid repr., model days with high generation cost at higher resolution. \\
    \textit{Method F (a posteriori)}: & Medoid repr., model days with high generation cost at higher resolution, cluster on time series inputs and storage patterns. \\ \hline
  \end{tabular}
  \caption{Time series aggregation schemes: three \textit{a priori} (A-C) and three \textit{a posteriori} (D-E).}
  \label{tab:methods}
\end{table}

\begin{table}
\centering
\small
\setlength{\tabcolsep}{0.3em}
\begin{tabular}{r c r l}
                                           & Number of  & \multicolumn{2}{r}{Solution time [min]} \\
  Aggregation                              & (repr.) days & Mean & (95\% of runs) \\ \hline
  \multicolumn{4}{c}{} \\
  \multicolumn{4}{c}{(a) \textbf{Validation}: 3-year base time series} \\ [0.6em]
  None (benchmark)                         & 1095$^*$   & 1005               & (322 - 3432)  \\
  A (\textit{a priori})                    & 30 \,      & 3                  & (2 - 6)       \\
  F (\textit{a posteriori})                & 30 \,      & 2+18+2 = 22        & (16 - 34)     \\
  A (\textit{a priori})                    & 120 \,     & 272                & (69 - 623)    \\
  F (\textit{a posteriori})                & 120 \,     & 299+19+165 = 483   & (133 - 887)   \\
  \multicolumn{4}{c}{} \\
  \multicolumn{4}{c}{(b) \textbf{Example}: 30-year base time series}  \\ [0.6em]
  None (benchmark)                         & 10950$^*$  & \textit{unfeasible}&  \\
  A (\textit{a priori})                    & 120 \,     & 974                & (97 - 1666)   \\
  F (\textit{a posteriori})                & 120 \,     & 968+301+717 = 1986 & (847 - 3770)  \\
  \multicolumn{4}{c}{} \\ \hline
  \multicolumn{4}{l}{$^*$ May include additional leap days}
\end{tabular}
  \caption{Simulations for (a) \textit{validation} (Section \ref{sec:simulations:validation}) and (b) \textit{example} (Section \ref{sec:simulations:example}) experiments. For simplicity, we detail one \textit{a priori} (A) and one \textit{a posteriori} method (F). For Method F, we disaggregate solution times into first planning, operation and second planning model runs (steps 1(b), 2(a) and 3 in Section \ref{sec:method:steps}).}
  \label{tab:simulations}
\end{table}

We examine six time series aggregation schemes as detailed in Table \ref{tab:methods}, all with daily periods. Methods A-C are \textit{a priori}. A and B use the cluster mean and medoid (closest real day to mean) respectively as representative day. Method C includes the maximum demand and minimum wind days in each region, a common \textit{a priori} way to preserve extremes (Section \ref{sec:introducion:time_series_aggregation}). Methods D-F are \textit{a posteriori}, using operational variables as described in Section \ref{sec:method:steps}. D and E include days with high unserved energy and generation cost, which serve as the \textit{importance} functions and correspond roughly to those used by \citet{Teichgraeber2021} and \citet{Hilbers2020} respectively in models without storage. Method F is the same as E, but uses storage (dis)charge decisions in the second aggregation (step 2(d)).

We use the following implementation. When we aggregate in steps 1(a) and 2(d), we scale and shift each time series to mean zero and variance one, reshape them to daily vectors and group them using Wald's hierarchical clustering. Representative days are either cluster means or medoids as specified in Table \ref{tab:methods}. For \textit{a posteriori} methods, we represent $p_e$=0.05 (5\%) of the original time series at higher resolution. We solve operational problems in step 2(a) sequentially using a horizon of one year and a window of six months. For example, we solve months 1-12 and store months 1-6, then solve 7-18 and store 7-12, etc.. To calculate the generation cost in Methods E and F, we assign a value of lost load of \pounds6,000/MWh \citep{elexon_2015} to unserved energy. We also run simulations with different choices than those presented here; these schemes showed similar or worse performance and are discussed in \ref{app:sims_not_used}.

We repeat experiments 40 times with different base time series $\myxiT$ created by resampling years with replacement. For example, a three-year sample may be [2011][1992][1992]. For reference, $\myDT$ has mean values (across 40 three-year time series) of 71.5GW baseload, 62.8GW peaking, 157.0GW wind, 153.3GW transmission and 427.9GWh storage. Our value of lost load implies $\approx$1\% additional system cost for every 0.01\% of energy not met.

\subsection{Validation}
\label{sec:simulations:validation}

\begin{figure}
  \centering
  \includegraphics[scale=0.8, trim=0 90 0 0, clip]{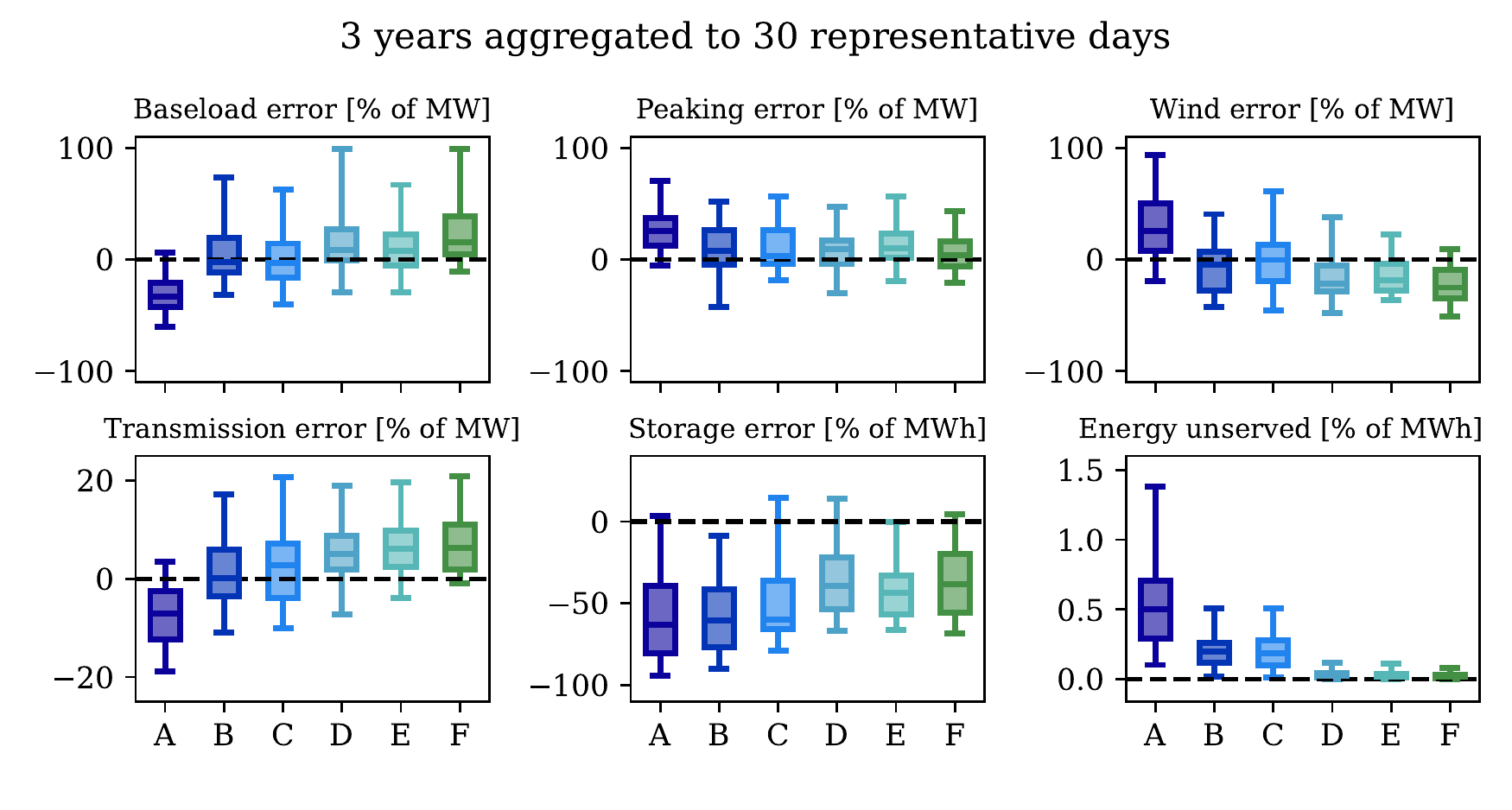}
  \includegraphics[scale=0.8, trim=0 0 0 0, clip]{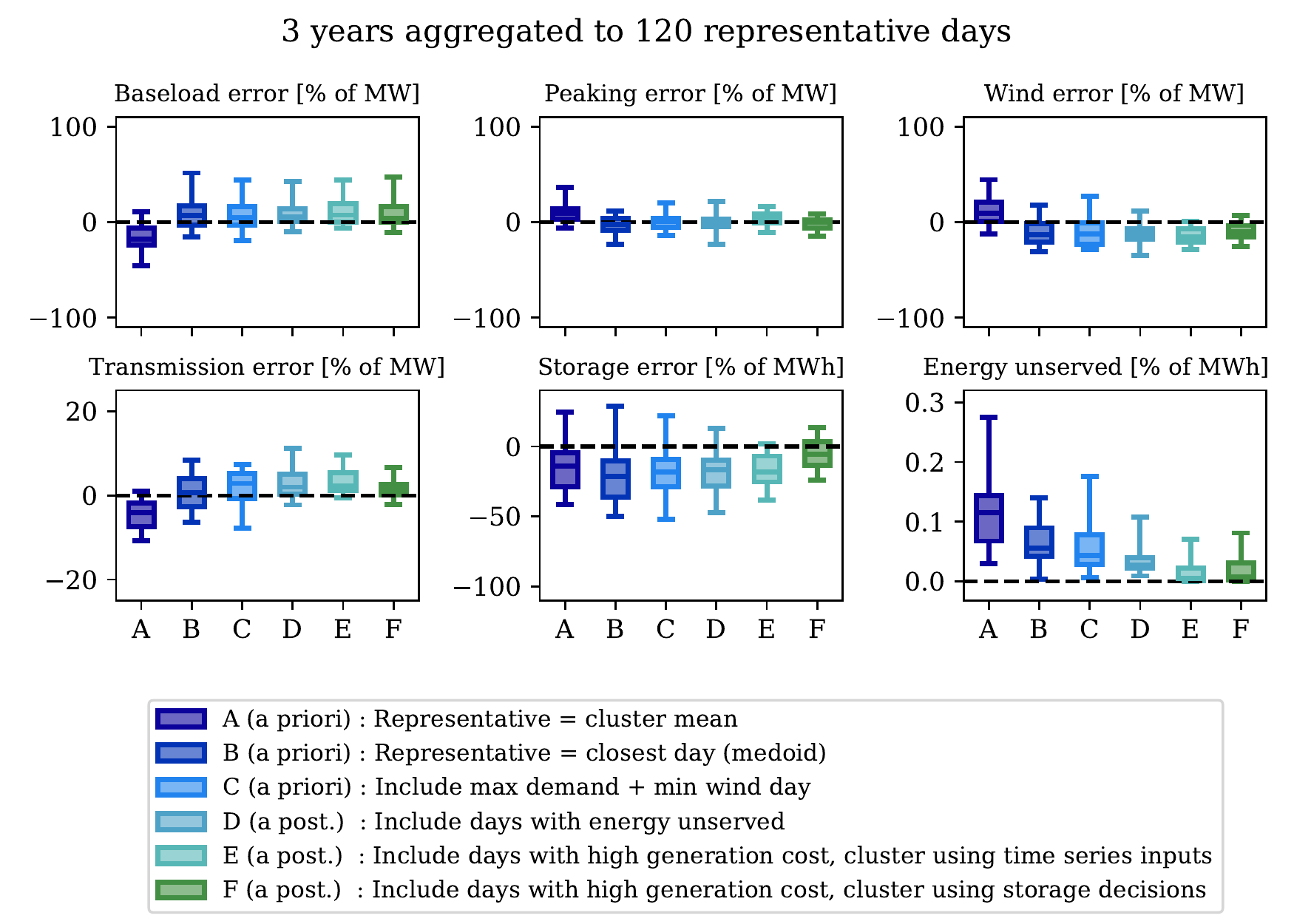}
  \caption{Distribution of evaluation metrics across 40 simulations for aggregation schemes A-F (Section \ref{sec:simulations:setup}, Table \ref{tab:methods}). The box and whiskers show the 2.5\%, 25\%, 50\%, 75\% and 97.5\% percentiles. We express values as percentages and denote the original unit, e.g.\ [\% of MW] is a percentage across values with unit MW.}
  \label{fig:results_validation}
\end{figure}

Figure \ref{fig:results_validation} shows results of the validation exercise on six metrics: percentage errors in baseload, peaking, wind, transmission and storage energy capacities (compared to the ``true'' optimum $\myDT$) as well as levels of unserved energy (MWh) across the full time series. For all methods, increasing the number of representative days from 30 to 120 decreases error metrics, but their relative performances differ.

Methods A-C are \textit{a priori}. Method A, with the cluster mean as representative day, overestimates peaking and wind while underestimating optimal baseload, transmission and (by a large margin) storage energy capacities, especially for 30 representative days. We observe unserved energy levels up to 1.5\% of demand. Method B, with medoid representative days, performs better; baseload, peaking, wind and transmission capacities are unbiased (median close to true value). Storage remains underestimated, but unserved energy is less than half of Method A. Method C, the \textit{a priori} attempt to include extremes via maximum demand and minimum wind days, does not further enhance performance.

The \textit{a posteriori} Methods D-F have significantly lower levels of unmet demand than A-C. Methods D and E, which identify extremes using unserved energy and generation cost respectively, show similar results. Method F, which uses storage (dis)charge decisions in clustering, more accurately estimates storage capacity.

Table \ref{tab:simulations}(a) shows solution times. For simplicity, we show \textit{a priori} Method A (B and C are similar) and \textit{a posteriori} Method F (D and E are similar). \textit{A posteriori} solution times consist of two planning runs across representative days and one operational run across the full time series without aggregation. While the operational solution times are constant, the two planning runs take much longer with 120 than 30 representative days.

\subsection{Example}
\label{sec:simulations:example}

\begin{figure}
  \centering
  \includegraphics[scale=0.8, trim=0 0 0 0, clip]{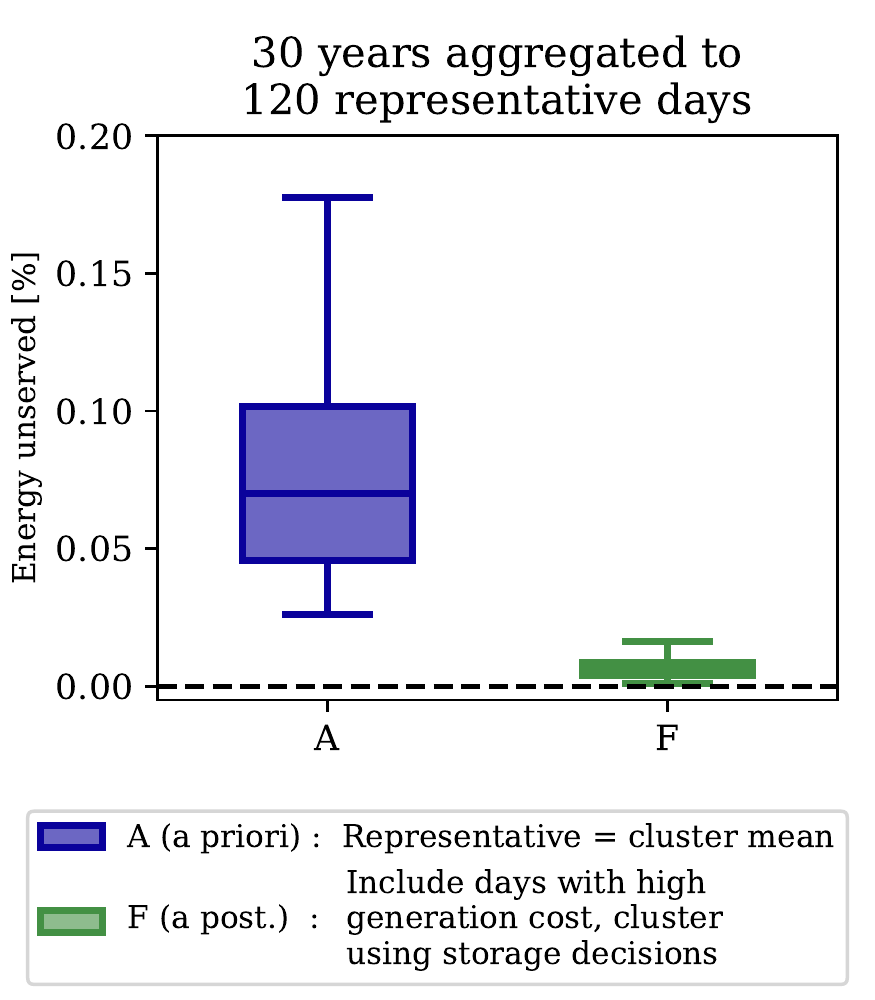}
  \caption{Distribution of unserved energy (as a percentage of total demand) across 40 simulations for aggregation methods A and F from 30 years to 120 representative days (Section \ref{sec:simulations:example}).}
  \label{fig:results_example}
\end{figure}

Figure \ref{fig:results_example} shows the results of aggregating 30 years into 120 representative days. In this case, the unaggregated problem is computationally unfeasible; extrapolating from shorter simulation lengths suggests over a year of solution time. We compare Method A (the most standard \textit{a priori} scheme) with F (the \textit{a posteriori} scheme with best performance). While unable to calculate capacity errors --- which require the non-aggregated design $\myDT$ --- unserved energy levels are around 20 times lower on average for Method F. This is a result of system designs that are more robust to extreme events; on average, designs for Method F have 30\% more baseload, 5\% less peaking, 22\% less wind, 11\% more transmission and 17\% more storage capacity than for Method A.

Table \ref{tab:simulations}(b) shows the distribution of solution times. Method F's solution times are formed mostly of the two planning runs on 120 representative days and not the operational run across 30 years.

\subsection{Discussion}
\label{sec:simulations:discussion}

For the \textit{a priori} methods, using the medoid (closest real day to cluster mean) as representative performs significantly better than using the mean itself. In fact, except in terms of storage energy capacity and unserved energy, it performs similarly to the more complicated \textit{a posteriori} methods. Our \textit{a priori} attempt at including extremes via the maximum demand and minimum wind days does not improve performance. As discussed in Section \ref{sec:introducion:time_series_aggregation}, this heuristic appears unable to determine those events that truly drive installed capacities.

The main performance enhancements from \textit{a posteriori} methods lie in reduced levels of unserved energy. This is to be expected, since they identify and include relevant extremes, ensuring resultant design estimates are robust to such occurrences. Our simulations suggest that both high generation cost and unserved energy successfully identify relevant extremes and can serve as useful \textit{importance} functions (Section \ref{sec:method:remarks}). The value in using storage (dis)charge decisions when aggregating (Method F) is concentrated primarily in more accurate estimates of optimal storage capacity, which many schemes underestimate significantly.

Obtaining a design estimate $\myDA{1}$ using an \textit{a posteriori} scheme requires two planning model runs (across $n_{\myA}$ representative periods each) and one operation model run (across the full time series, length $n_{\myT}$ periods). Both the accuracy of design estimates and computational times increase with the number of representative periods. Hence, in practice, the largest number of representative periods that are computationally feasible should be chosen; in this case, the two planning runs usually constitute the majority of solution times.

\section{Conclusions}
\label{sec:conclusion_implications}

\subsection{Conclusions}
\label{sec:conclusion_implications:conclusion}

This paper introduces a framework for \textit{a posteriori} time series aggregation schemes for energy system (capacity expansion) planning models with storage. They allow us to estimate optimal system design or investment decisions across long time series at significantly reduced computational cost and with smaller levels of error than established approaches. Our models, data and code are publicly available at \doi{10.5281/zenodo.7178301}.

The ability to reliably consider long samples in capacity expansion planning problems has two important implications. Firstly, it reduces the impact of sampling uncertainty on model outputs, reducing the risk in incorrect strategical decisions as a result of an unrepresentative demand and weather time series. Secondly, it is a step towards the use of climate model data --- typically many multi-decadal samples from an ensemble of simulations --- in planning models: without reliable compression techniques the computation is unfeasible.

Our framework customises aggregation to the energy system model using its operational variables (generation, transmission and storage patterns). It unifies methods by \citet{Hilbers2020} and \citet{Teichgraeber2021} --- which use operational variables in models without storage --- with that of \citet{Kotzur2018a} --- which allow chronology-preserving aggregation for storage technologies.

Our experiments motivate a number of recommendations. In line with previous studies in Section \ref{sec:introducion:time_series_aggregation}, medoid-based aggregation should be considered as an \textit{a priori} adjustment (no increase in computational cost) that can significantly improve performance. Aggregation should preserve relevant extreme events (usually those which require peak generation capacities). These can be identified \textit{a posteriori} using operational variables; we find that \textit{a priori} heuristics such as including the maximum demand or minimum renewable generation days may not work well.

\subsection{Extensions}
\label{sec:conclusion_implications:implications}

Our framework uses two planning runs to obtain $\myDA{0}$ and $\myDA{1}$. We can, however, iterate steps 2 and 3, repeatedly using the last design estimate to calculate new operational variables and design estimates $\myD_{\myA_i}$ for $i > 1$. This is done by \cite{Bahl2018} and \cite{Teichgraeber2021} with unserved energy as the \textit{importance} function in models without storage. In our experiments, we find further iterations to offer minimal performance gain, but this need not hold in general.

Another extension involves dimensionality reduction. Even in our comparatively simple case study models, our three time series of demand, wind generation potentials and storage (dis)charge decisions give 3$\times$3$\times$24 = 216 components in each daily vector to cluster in steps 1(a) and 2(d) (Section \ref{sec:method:steps}). For models with more time series inputs, reducing dimensionality may be necessary.

We may also use more information from operational variables. In our case studies, we use (1) generation cost or unserved energy levels to identify extreme events and (2) storage patterns when clustering. However, the operational variables include more information, such as generation levels of individual technologies and regions, from which we may be able to extract more information. We may also investigate different new \textit{importance} functions, such as the electricity price.

Other classes of extensions involve combining different chronology-preserving or \textit{a posteriori} aggregation. For example, we may use the \textit{system states} or \textit{chronological time period clustering} (Section \ref{sec:introduction:inter_period_links}) to link periods across time, or cluster in solution space as some methods in Section \ref{sec:introduction:adaptive_methods}.

Finally, we can reduce our models' operational foresight. For our planning and operational models, we optimise operation across the full sample and one year ahead respectively. Planning problems without perfect foresight are more realistic but may require different solution methods.

\section*{Acknowledgements}

This work was supported by the Engineering and Physical Sciences Research Council (EPSRC) Mathematics of Planet Earth Centre for Doctoral Training, grant number EP/L016613/1.

\appendix
\section{Simulations not included}
\label{app:sims_not_used}

A number of experiments showed performance that was either similar or worse than those presented in Section \ref{sec:siss:simulations}. These included using $p_e$=0.1 instead of 0.05, meaning 10\% of the time series was considered ``extreme''. We also used different vector normalisation, including scaling and shifting time series to lie between zero and one (instead of having mean zero and variance one) and normalising \textit{daily} vectors (so that each hour-of-day in each time series has mean zero and variance one instead of each time series). To determine extreme periods, we included days with peak generation cost or unserved energy instead of maximum integral value. Finally, we calculated additional design estimates $\myDA{i}$ for $i>1$ by repeating steps 2 and 3 in Section \ref{sec:method:steps}.

\section{Planning models: mathematical details}
\label{app:psm}

\begin{table}
  \noindent
  \begin{tabular}[]{r l }
    \multicolumn{2}{l}{\textit{Indices \& Sets}} \\
    \hspace*{1.0em} $i \in \mathcal{I}$ & Generation technology \\
    $r \in \mathcal{R}$ & Region \\
    $t \in \mathcal{T}$ & Time step \\
    \multicolumn{2}{l}{\textit{Parameters}} \\
    $C_i^\gen $ & Annualised generation install cost, technology $i$ [\pounds /MWyr] \\
    $C_{r, r'}^\tr $ & Annualised transmission install cost, region $r$ to $r'$ [\pounds /MWyr] \\
    $C^\sto $ & Annualised storage energy install cost [\pounds /MWhyr] \\
    $F_i$ & Generation cost, technology $i$ [\pounds /MWh] \\
    $e^\sto$ & Storage (dis)charge efficiency [$\in$ [0, 1]] \\
    $l^\sto$ & Storage self-loss [1/hr] \\
    \multicolumn{2}{l}{\textit{Time series}} \\
    $d_{r, t}$ & Demand, region $r$, time $t$ [MWh] \\
    $\lambda_{i, r, t}$ & Generation potential, technology $i$, region $r$, time $t$ [$\in$ [0, 1]] \\
    $\myxi_t$ & Time series values, time $t$ \\
    \multicolumn{2}{l}{\textit{Decision variables}} \\
    $\capgen_{i, r}$ & Generation capacity, technology $i$, region $r$ [MW] \\
    $\captrans_{r, r'}$ & Transmission capacity, region $r$ to $r'$ [MW] \\
    $\capsto_r$ & Storage energy capacity, region $r$ [MWh] \\
    $\gen_{i, r, t}$ & Generation, technology $i$, region $r$, time $t$ [MWh] \\
    $\tr_{r, r', t}$ & Transmission, region $r$ to $r'$, time $t$ [MWh] \\
    $\ch_{r, t}$ & Storage charging, region $r$, time $t$ [MWh] \\
    $\sto_{r, t}$ & Storage energy level, region $r$, time $t$ [MWh] \\
    $\myD$ & Power system design \\
    $\myO_t$ & Power system operation, time $t$
  \end{tabular}
  \caption{Nomenclature.}
  \label{table:nomenclature}
\end{table}

\begin{table}
  \centering
  \small
  \setlength{\tabcolsep}{0.3em}
  \begin{tabular}{ l  c  c  c  c  c  c}
                     & \multicolumn{3}{c}{Monetary cost}                 & \multicolumn{2}{c}{Storage}  \\
                     & Install        & Install         & Generation     & Efficiency & Self-loss       \\
    Technology ($i$) & [\pounds/MWyr] & [\pounds/MWhyr] & [\pounds /MWh] & [1]         & [1/hr]         \\ \hline
                     &                &                 &                &            &                 \\
    \textit{Generation} & $C^\gen_i$   &                 & $F_i$          &            &                 \\ \hline
    Baseload ($b$)   & 300,000        & ---             & 5              & ---        & ---             \\
    Peaking ($p$)    & 100,000        & ---             & 35             & ---        & ---             \\
    Wind ($w$)       & 100,000        & ---             & ---            & ---        & ---             \\ \hline
                     &                &                 &                &            &                 \\
    \textit{Transmission} & $C^{\tr}_{r,r'}$ &            &                &            &                 \\ \hline
    Region 1-5       & 150,000        & ---             & ---            & ---        & ---             \\
    Other            & 100,000        & ---             & ---            & ---        & ---             \\
                     &                &                 &                &            &                 \\
    \textit{Storage} &                & $C^{\sto}$       &                & $e^\sto$    & $l^\sto$        \\ \hline
    Storage          & ---            & 1,000           & ---            & 0.95       & 0.00001         \\ \hline
  \end{tabular}
  \caption{Technologies. Install costs are expressed per year of infrastructure lifetime. Carbon emissions are expressed in kg CO$_2$ equivalent warming potential. Storage efficiency is dimensionless. To avoid solution nonuniqueness, costs are perturbed slightly ($<0.1\%$) in different regions.}
  \label{table:technologies}
\end{table}

Our planning model's generation, transmission and storage technologies are detailed in Table \ref{table:technologies}. Time series inputs (hourly demand levels in Regions 2, 4 and 5; wind \textit{generation potentials} (capacity factors) in Regions 2,5 and 6) contain data across Europe for 1980-2017, as introduced by \citet{bloomfield_2019} and available at \citep{bloomfield_MERRA2}. Let $\mathcal{I}$=$\{b, p, w\}$ and $\mathcal{R}$=$\{1, 2, 3, 4, 5, 6\}$ be the generation technologies (baseload, peaking, wind) and regions respectively. Then $\myxi_t = [d_{2,t}, d_{4,t}, d_{5,t}, \lambda_{w,2,t}, \lambda_{w,5,t}, \lambda_{w,6,t}]$ is the demand and weather data at time $t$. The planning problem is to minimise
\begin{equation}
\sum_{r \in \mathcal{R}} \Bigg[ \frac{T}{8760}
  \Bigg( \! \underbrace{\sum_{i \in \mathcal{I}} C_i^\gen \capgen_{i,r}}_{\substack{\text{install cost,} \\ \text{generation}}}
  + \!\! \underbrace{ \sum_{r' \in \mathcal{R}} C_{r,r'}^\tr \captrans_{r,r'} }_{\substack{\text{install cost,} \\ \text{transmission}}}
  + \underbrace{ \vphantom{\sum_{i \in \mathcal{I}}} C^\sto \capsto_r }_{\substack{\text{install cost,} \\ \text{storage}}} \!\! \Bigg)
  + \underbrace{ \sum_{t \in \mathcal{T}} \sum_{i \in \mathcal{I}} F_i^\gen \gen_{i,r,t}}_\text{generation cost} \Bigg]
\label{eq:test_psm:plan:objective} \vspace*{0.5em}
\end{equation}
by optimising over design $\myD$ and operation $(\myO_t)_{t \in \myT}$, where
\begin{align}
  \myD = [\capgen_{i,r}, \ \captrans_{r,r'}, \ \capsto_r \ | \ i \in \mathcal{I}; \ r \in \mathcal{R}; \ r' \in \mathcal{R}] \\
  \myO_t = [\gen_{i,r,t}, \ \tr_{r,r',t}, \ \ch_{r, t} \ | \ i \in \mathcal{I}; \ r \in \mathcal{R}; \ r' \in \mathcal{R}]
\label{eq:test_psm:plan:decision_variables}
\end{align}
\noindent subject to
\begin{align}
  \capgen_{b, r} \big\rvert_{r \notin \{1,3,6\}} = \capgen_{p,r} \big\rvert_{r \notin \{1,3,6\}} = \capgen_{w, r} \big\rvert_{r \notin \{2,5,6\}} &= 0 \label{eq:test_psm:plan:topology_gen} \\
  \captrans_{r,r'} \big\rvert_{(r,r') \notin  \{(1,2), (1,5), (1,6), (2,3), (3,4), (4,5), (5,6)\}} &= 0 \label{eq:test_psm:plan:topology_tr} \\
  \capsto_r \big\rvert_{r \notin \{2,5,6\}} &= 0 \label{eq:test_psm:plan:topology_sto} \\
  \sum_{i \in \mathcal{I}} \gen_{i,r,t} + \sum_{r' \in \mathcal{R}} \tr_{r',r,t} = d_{r,t} + \ch_{r,t} \quad & \text{for all} \ r, t \label{eq:test_psm:operate:demand_met} \\
  \tr_{r,r',t} + \tr_{r,'r,t} = 0 \quad & \text{for all} \ r, r', t \label{eq:test_psm:operate:tr_balance} \\
  \sto_{r,0} = 0 \quad & \text{for all} \ r \label{eq:test_psm:operate:storage_initial} \\
  \sto_{r,t+1} = (1 - l^\text{sto}) \sto_{r,t} +
  \begin{cases}
    e^\sto \ch_{r,t} \quad \text{if} \ \ch_{r,t} \ge 0 \\
    \frac{1}{e^\sto} \ch_{r,t} \quad \text{if} \ \ch_{r,t} < 0
  \end{cases}
  & \text{for all} \ r, t \label{eq:test_psm:operate:storage_continuity} \\
  0 \le \gen_{i,r,t} \le \capgen_{i,r} \quad & \text{for all} \ i, r, t \label{eq:test_psm:operate:gen_le_cap_conv} \\
  0 \le \gen_{w,r,t} \le \capgen_{w,r} \lambda_{w,r,t} \quad & \text{for all} \ r, t \label{eq:test_psm:operate:gen_le_cap_wind} \\
  |\tr_{r,r',t}| \le \captrans_{r,r'} + \captrans_{r',r} \quad & \text{for all} \ r, r', t \label{eq:test_psm:operate:tr_le_cap_tr} \\
  0 \le \sto_{r,t} \le \capsto_r \quad & \text{for all} \ r, t \label{eq:test_psm:operate:sto_within_bounds}.
\end{align}
For definitions of terms and parameter values, see Tables \ref{table:nomenclature} and \ref{table:technologies} respectively. The factor $\frac{T}{8760}$ normalises install costs to the same temporal scale as generation costs, since $C_i^{gen}$ and $C_{r,r'}^{tr}$ are costs per year of plant lifetime and there are 8760 hours (time steps) in a year.

The constraints have the following meanings.
\eqref{eq:test_psm:plan:topology_gen}-\eqref{eq:test_psm:plan:topology_sto} are the model's generation, transmission and storage topology.
\eqref{eq:test_psm:operate:demand_met} indicates that generation plus transmission into a region equals demand plus storage charging.
\eqref{eq:test_psm:operate:tr_balance} is the transmission balance.
\eqref{eq:test_psm:operate:storage_initial} specifies empty initial storage and
\eqref{eq:test_psm:operate:storage_continuity} indicates how storage levels change with storage (dis)charging and self-loss.
\eqref{eq:test_psm:operate:gen_le_cap_conv}-\eqref{eq:test_psm:operate:gen_le_cap_wind} ensure generation does not exceed installed capacity (for thermal technologies) or installed capacity times generation potential (for wind).
\eqref{eq:test_psm:operate:tr_le_cap_tr} limits transmitted power to installed transmission capacity.
\eqref{eq:test_psm:operate:sto_within_bounds} constraints storage levels to lie within storage (energy) bounds.

\section*{Bibliography}
\label{sec:bib}

\small{\bibliography{../../citations/citations}}

\end{document}